\documentclass[twoside]{ilcws07}
\usepackage[utf8]{inputenc}
\usepackage[dvips]{graphicx,epsfig,color}
\usepackage{wrapfig,rotating}
\usepackage{amssymb,amsmath,array}

\def\reffi#1{Figure~\ref{#1}}
\def\citere#1{Ref.~\cite{#1}}

\newcommand{\Pep}{\mathrm {e^+}}
\newcommand{\Pem}{\mathrm {e^-}}
\newcommand{\PH}{\mathrm H}
\newcommand{\mh}{\ensuremath{m_\mathrm{H}}}

\newcommand{\GeV}{\ensuremath{\,\text{GeV}}}

\newcommand{\fb}{\ensuremath{\,\text{fb}}}

\newcommand{\order}[1]{\ensuremath{{\cal O}(#1)}}

\newcommand{\eennh}{\ensuremath{\Pep\Pem\to\nu\bar\nu\PH}}
\newcommand{\qqqqh}{\ensuremath{qq' \to qq'H}}

\newcommand{\ggqqh}{\ensuremath{gg \to q\Bar{q}H}}
\newcommand{\qqggh}{\ensuremath{q\Bar{q} \to ggH}}
\newcommand{\qgqgh}{\ensuremath{qg \to qgH}}
\newcommand{\qbgqbgh}{\ensuremath{\Bar{q}g \to \Bar{q}gH}}

\begin{document}


\title{
Higgs Production by Gluon initiated Weak Boson Fusion} 
\author{M. M. Weber$^1$
\vspace{.3cm}\\
1- Department of Physics, University at Buffalo \\
The State University of New York, Buffalo, NY 14260-1500, USA
}

\maketitle

\begin{abstract}
  The gluon-gluon induced terms for Higgs production through
  weak-boson fusion are calculated. They form a finite and
  gauge-invariant subset of the NNLO corrections in the strong
  coupling constant. This is also the lowest order with sizeable
  t-channel colour exchange contributions, leading to additional
  hadronic activity between the outgoing jets.
\end{abstract}



\section{Introduction}

The weak-boson-fusion (WBF) process $\qqqqh$ is one of the major
Higgs-boson production processes at the LHC. With a cross section of
up to 20\% of the leading gluon fusion process for low Higgs masses it
allows a discovery of the Higgs boson in the intermediate mass range
as well as for high masses \cite{Asai:2004ws, Abdullin:2005yn}.
Furthermore it allows for precise measurements of the Higgs couplings.

Weak-boson fusion has a characteristic signature that can be used to
separate it well from the background processes
\cite{Rainwater:1996ud}. Since the LO diagrams do not contain
t-channel exchange of coloured particles the final-state quarks appear
as jets in opposite hemispheres at high rapidities.  In the central
region between the jets there is very little hadronic activity, only
the Higgs decay products are found here.

At NLO the QCD corrections to total rates
\cite{Han:1992hr,Djouadi:1999ht} and the differential cross section
\cite{Figy:2003nv,Berger:2004pc} have been calculated. They increase
the cross section at the LHC by about 10\% while reducing the residual
scale dependence to about 3\%.  Colour exchange contributions are
strongly suppressed at NLO since diagrams with t-channel gluon
exchange contribute only through their interference with u-channel
Born diagrams. Since the interference between t- and u-channel
diagrams is very small it is usually neglected.  In this approximation
there are no colour exchange contributions even at NLO and the
corrections can be expressed in terms of the structure functions of
deep inelastic scattering.  Recently also the NLO electroweak
corrections and the QCD corrections to the interference terms and have
been calculated \cite{Ciccolini:2007jr}.

Contributions with sizeable t-channel colour exchange can first appear
at NNLO.  Although at this order gluon or quark pairs can be exchanged
between the quark lines, the non-colour-singlet part contributes only
in the interference with u-channel diagrams and can therefore be
expected to be tiny.  Another part of the NNLO corrections is the
square of the $\order{\alpha_s}$ amplitudes. In this contribution
non-suppressed diagrams with net colour exchange may appear leading to
a possible deviation from the characteristic signature of WBF.
Furthermore the NNLO corrections might be larger than could be
expected from a naive extrapolation of the NNLO DIS results.

In order to assess the size of these effects we have studied the
process $\ggqqh$ and the crossed processes $\qqggh$, $\qgqgh$ and
$\qbgqbgh$ \cite{Harlander-wbf}. The amplitude is of
$\order{\alpha_s}$ and its square therefore contributes to WBF at
NNLO. Since these are loop induced processes appearing first in this
order, they are a UV-finite and gauge-invariant subset of the full
NNLO corrections. Due to the large gluon luminosity at the LHC they
can also be expected to constitute a sizeable part of the complete
NNLO corrections.

The same process also appears in the real-emission corrections to
Higgs production in gluon fusion at NNLO.  The production of the
resulting H+2jet final states has been studied in
\cite{DelDuca:2001fn}.  However, since these diagrams are of a
different order in the coupling constant we will not consider them in
this work.



\section{Calculational Framework}

\begin{figure}[t]
\centerline{\begin{tabular}{ccc}
{\includegraphics[scale=0.3]{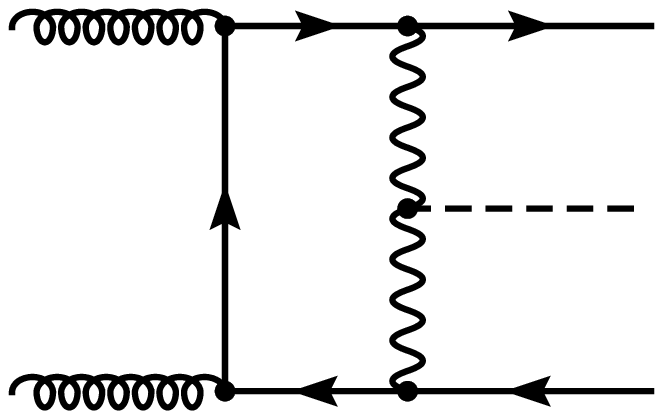}}&
{\includegraphics[scale=0.3]{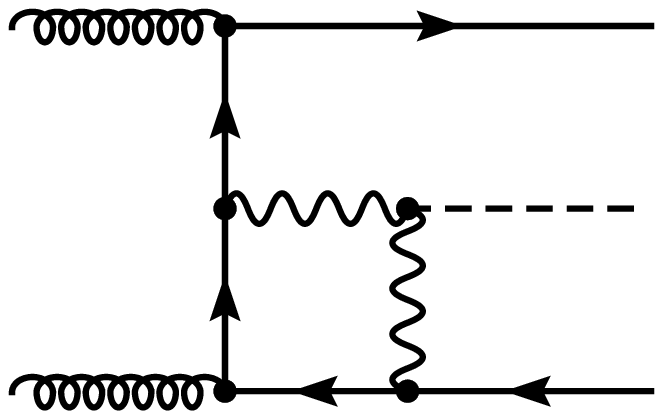}}&
{\includegraphics[scale=0.3]{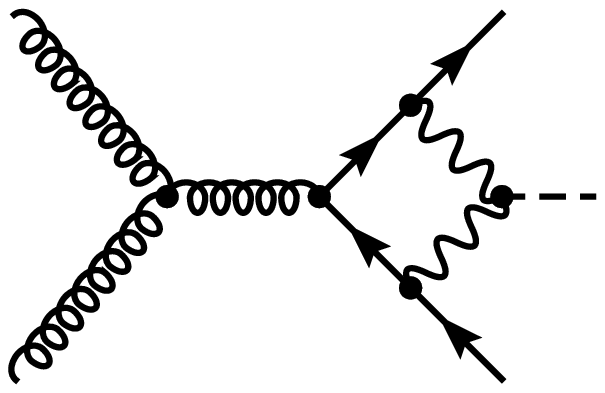}}\\
{\includegraphics[scale=0.3]{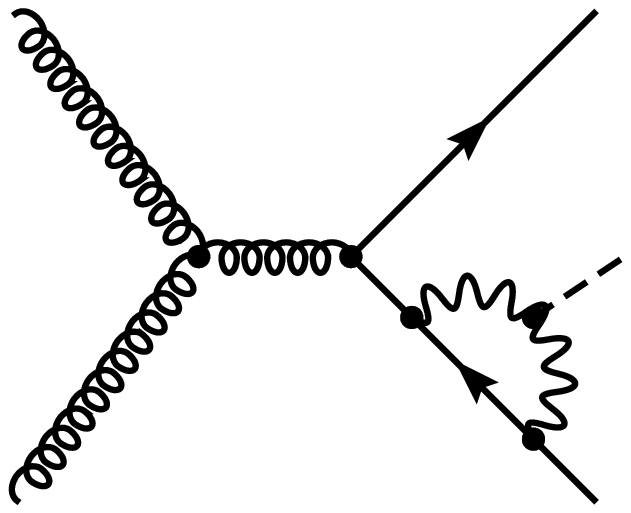}}&
{\includegraphics[scale=0.3]{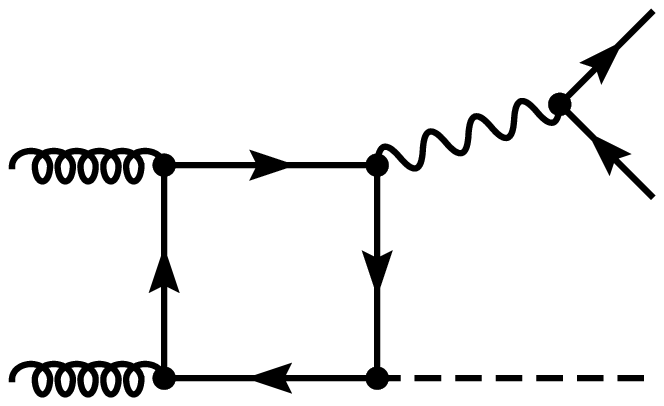}}&
{\includegraphics[scale=0.3]{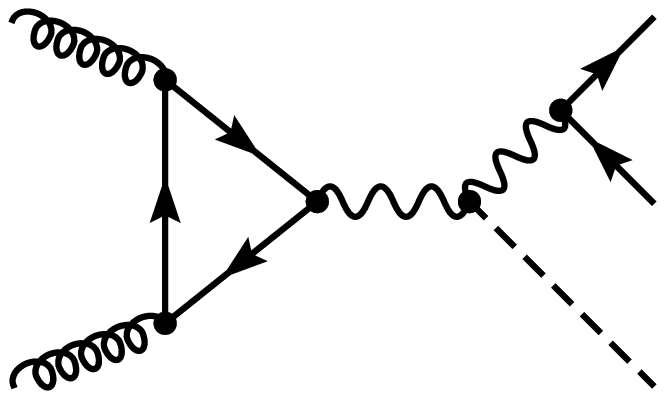}}
\end{tabular}}
\caption{Sample diagrams for the process $\ggqqh$.}
\label{fig:diags}
\end{figure}

An overwiev of the calculation using the process $\ggqqh$ as an
example is given in the following. A full description of the complete
calculation can be found in \citere{Harlander-wbf}.  Some sample
1-loop diagrams are shown in \reffi{fig:diags}.  We treat the quarks
including the b-quark as massless, and always sum over all 5 light
flavours. With this approximation the Higgs boson couples only to the
weak gauge bosons and to closed top-quark loops.

The last two diagrams belong to a class containing a virtual $Z$-boson
splitting into a final state $q\Bar{q}$ pair.  These diagrams form a
gauge-invariant subset.  The $Z$-boson may become resonant and this
class then describes $HZ$ production with a subsequent $Z \to
q\Bar{q}$ decay. Consequently these diagrams belong to the NNLO
corrections to the Higgsstrahlung process $q\Bar{q} \to HZ$ and have
to be taken into account there.  Since we are only interested in WBF
this diagram class is discarded in the following.

Furthermore some diagrams are part of real corrections to lower order
Higgs-production processes. Since these are singular in the soft and
collinear parts of the phase space we require the two final-state
quarks to form two well-resolved jets. With this restriction all
diagrams are IR finite over the whole remaining phase space and one
obtains a well-defined total rate.

Technically the most challenging part of the calculation are the
5-point diagrams like the first one in \reffi{fig:diags}.  These
diagrams are similar to the ones appearing in the recent calculation
of the electroweak corrections to the process $\eennh$
\cite{Denner:2003yg} and the same techniques can also be applied in
this case.

The actual calculation of the diagrams has been performed using the 't
Hooft--Feynman gauge. The graphs were generated by {\sl FeynArts}
\cite{Hahn:2000kx} and the evaluation of the amplitudes performed
using {\sl FormCalc} \cite{Hahn:1998yk}. The analytical results of
{\sl FormCalc} in terms of Weyl-spinor chains and coefficients
containing the tensor loop integrals have been translated to {\sl C++}
code for the numerical evaluation.  The tensor and scalar 5-point
functions are reduced to 4-point functions following
\citere{Denner:2002ii}, where a method for a direct reduction is
described that avoids leading inverse Gram determinants which can
cause numerical instabilities.  The remaining tensor coefficients of
the one-loop integrals are recursively reduced to scalar integrals
with the Passarino--Veltman algorithm \cite{Passarino:1979jh} for
non-exceptional phase-space points.  In the exceptional phase-space
regions the reduction of the 3- and 4-point tensor integrals is
performed using the methods of \citere{Denner:2005nn} which allow for
a numerically stable evaluation.

The phase-space integration is performed with Monte Carlo techniques
using the adaptive multi-dimensional integration program {\sc Vegas}
\cite{Lepage:1977sw}.



\section{Numerical Results}

To study the impact of the contribution calculated here we compare the
total cross section to the LO result for WBF.  In order to get a well
defined total rate we always employ a minimal set of cuts.  These
minimal cuts ensure two well-separated jets in the final state and are
given by
\[ p_{T_j} > 20 \GeV, \quad |\eta_j|<5, \quad R>0.6,\]
where $p_{T_j}$ and $\eta_j$ are the transverse momenta and the
pseudorapidities of the final state jets emerging from the quarks and
gluons and
\[ R = \sqrt{(\Delta\eta)^2 + (\Delta\phi)^2}\]
with $\Delta\eta = \eta_1 - \eta_2$ and $\Delta\phi = \phi_1 - \phi_2$
is the separation of the jets in the pseudorapidity--azimuthal angle
plane.

A much improved signal-to-background ratio for weak-boson fusion can
be obtained by further cuts \cite{Rainwater:1996ud}.  These additional
WBF cuts require that the two jets are well separated, reside in
opposite detector hemispheres and have a large dijet invariant mass
\[ |\Delta\eta| > 4.2, \quad \eta_1 \cdot \eta_2 < 0, \quad m_{jj} > 600 \GeV. \]

\begin{figure}[t]
\centerline{
\includegraphics[scale=0.95]{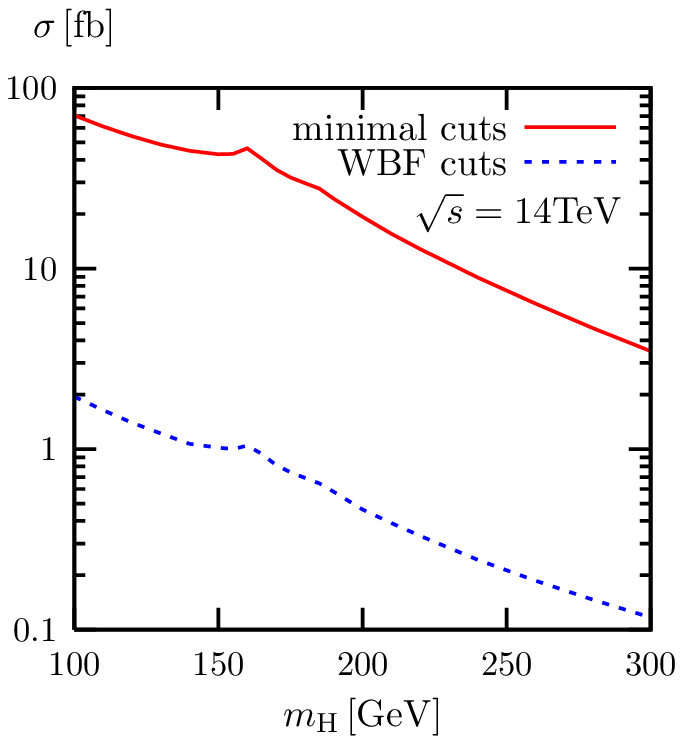}
\qquad
\includegraphics[scale=0.95]{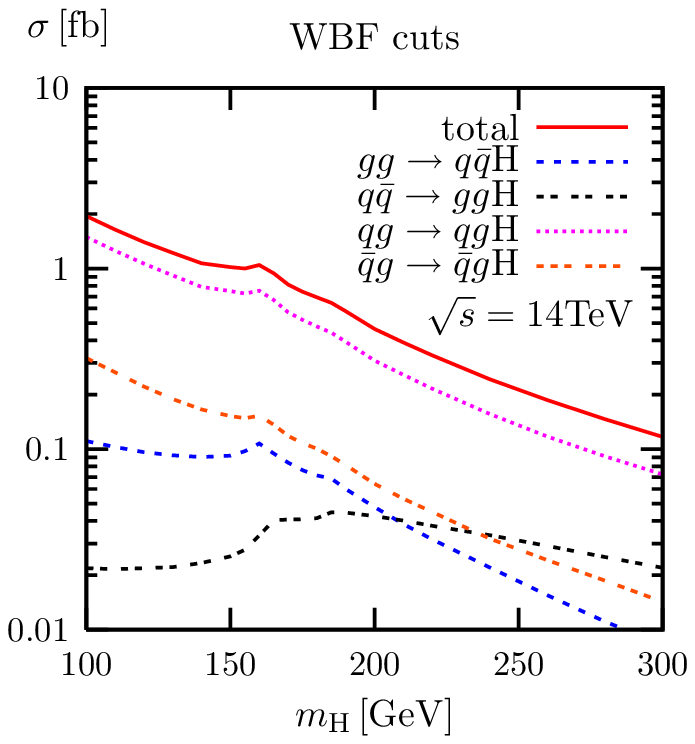}
}
\caption{Total cross section summed over all processes with minimal cuts and WBF cuts applied (l.h.s) and separate contributions of all processes using WBF cuts (r.h.s).}
\label{fig:mhplot}
\end{figure}

The total cross section summed over all crossed processes is shown on
the l.h.s of \reffi{fig:mhplot} as a function of the Higgs-boson mass.
With only the minimal cuts employed the total rate is about $70 \fb$
for low Higgs masses and falls off steeply towards higher masses. At
$\mh=100\GeV$ this amounts to about $2\%$ of the LO cross section for
WBF which is in accordance with the naive expectation for the order of
magnitude of the NNLO corrections. The decrease of the cross section
toward higher $\mh$ is however much steeper for the $\ggqqh$ process
than the rather moderate decrease of the LO result.

The effect of the additional WBF cuts is a strong suppression of the
cross section by roughly a factor $30$. This strong suppression is in
contrast to the LO and NLO WBF rates which only show a suppression by
about a factor of $2 - 3$. As the WBF cuts are designed to take advantage
of the characteristic signature of weak-boson fusion, this indicates
that the kinematics of the contribution investigated here is rather
different than the normal WBF kinematics.

The cross sections for the separate processes using WBF cuts are shown
on the r.h.s of \reffi{fig:mhplot}. The largest contribution comes
from the process $\qgqgh$ while all other processes are at least a
factor of 3 smaller.

\begin{figure*}[t]
\centerline{
\includegraphics[scale=0.95]{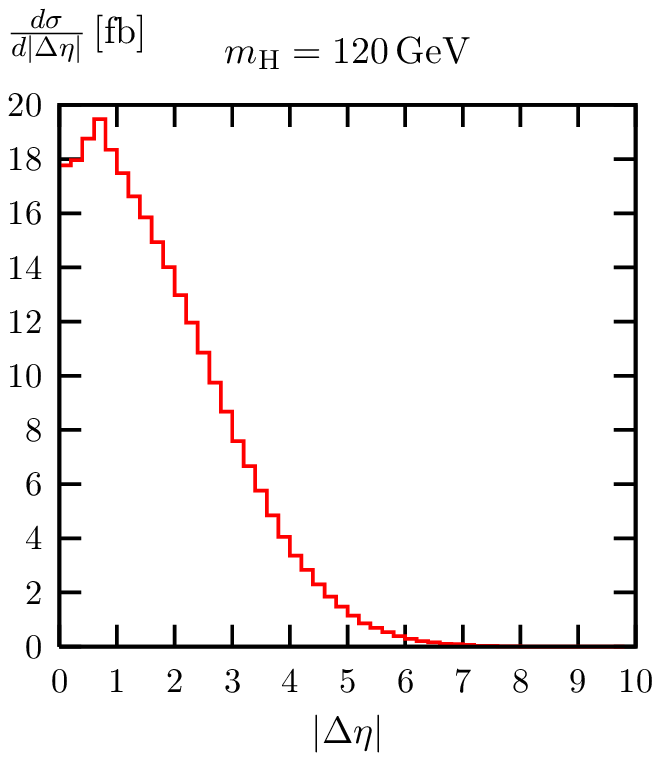}
\qquad
\includegraphics[scale=0.95]{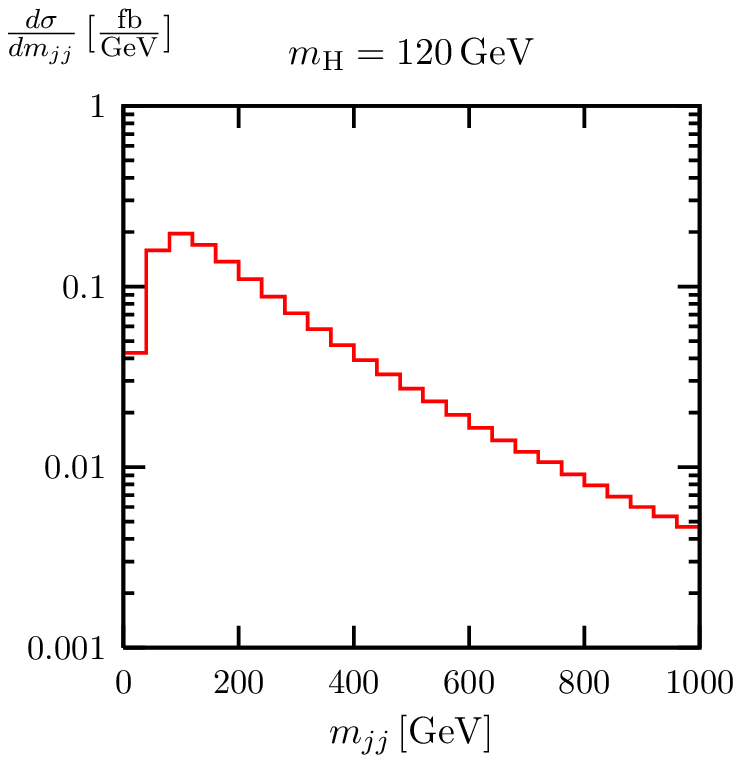}
}
\caption{Distribution in the pseudorapidity separation of the quark jets
(l.h.s) and the jet-jet invariant mass (r.h.s) for a Higgs mass $m_\PH
= 120 \GeV$.}
\label{fig:histo}
\end{figure*}

In order to shed more light on the origin of the strong suppression
caused by the WBF cuts the behaviour of the quantities appearing in
the cuts has to be investigated. Therefore the distributions in the
pseudorapidity separation $\Delta\eta$ and the jet-jet invariant mass
$m_{jj}$ are shown in \reffi{fig:histo} for a Higgs mass of
$\mh=120\GeV$. The pseudorapidity-separation is peaked at low values
of about 1 which is much lower than the corresponding peak for the LO
result located at $\Delta\eta \simeq 4$
\cite{Figy:2003nv,Berger:2004pc}. The invariant mass distribution
falls off fast with increasing $m_{jj}$. This falloff is stronger than
for WBF at leading order. This shows that the jets are less well
separated than for the LO WBF and therefore suffer a stronger
suppression by the additional WBF cuts.

\section{Summary}

We have performed a calculation of the loop-induced process $\ggqqh$
and the crossed processes.  These are a gauge-invariant and finite
part of the NNLO corrections to weak-boson fusion featuring t-channel
colour exchange, which is strongly suppressed at lower orders. The
total cross section is about $70\fb$ at $\mh=100\GeV$ and falls off
towards higher Higgs masses. Imposing further cuts commonly used to
separate the weak-boson-fusion signal from background leads to a
strong suppression of the total rates by about a factor of $30$. An
investigation of distributions has shown this to be caused by
different kinematics than for the leading-order weak-boson-fusion
process.

\section*{Acknowledgements}

We thank Ansgar Denner for supplying us with his Fortran library for the
evaluation of the scalar and tensor loop integrals.




\begin{footnotesize}

\end{footnotesize}



\begin{thebibliography}{99}
\bibitem{url} Slides: \\
\verb$http://ilcagenda.linearcollider.org/contributionDisplay.py?contribId=409&sessionId=73&confId=1296$

\bibitem{Asai:2004ws}
  S.~Asai {\it et al.},
  Eur.\ Phys.\ J.\  C {\bf 32S2} (2004) 19
  [arXiv:hep-ph/0402254].

\bibitem{Abdullin:2005yn}
  S.~Abdullin {\it et al.},
  Eur.\ Phys.\ J.\  C {\bf 39S2} (2005) 41.

\bibitem{Rainwater:1996ud}
  D.~L.~Rainwater, R.~Szalapski and D.~Zeppenfeld,
  Phys.\ Rev.\ D {\bf 54} (1996) 6680
  [arXiv:hep-ph/9605444];
  D.~L.~Rainwater and D.~Zeppenfeld,
  Phys.\ Rev.\ D {\bf 60} (1999) 113004
  [Erratum-ibid.\ D {\bf 61} (2000) 099901]
  [arXiv:hep-ph/9906218];
  T.~Plehn, D.~L.~Rainwater and D.~Zeppenfeld,
  Phys.\ Rev.\ D {\bf 61} (2000) 093005
  [arXiv:hep-ph/9911385].

\bibitem{Han:1992hr}
  T.~Han, G.~Valencia and S.~Willenbrock,
  Phys.\ Rev.\ Lett.\  {\bf 69} (1992) 3274
  [arXiv:hep-ph/9206246].

\bibitem{Djouadi:1999ht}
  A.~Djouadi and M.~Spira,
  Phys.\ Rev.\ D {\bf 62} (2000) 014004
  [arXiv:hep-ph/9912476].

\bibitem{Figy:2003nv}
  T.~Figy, C.~Oleari and D.~Zeppenfeld,
  Phys.\ Rev.\ D {\bf 68} (2003) 073005
  [arXiv:hep-ph/0306109].

\bibitem{Berger:2004pc}
  E.~L.~Berger and J.~Campbell,
  Phys.\ Rev.\ D {\bf 70} (2004) 073011
  [arXiv:hep-ph/0403194].

\bibitem{Ciccolini:2007jr}
  M.~Ciccolini, A.~Denner and S.~Dittmaier,
  arXiv:0707.0381 [hep-ph].

\bibitem{Harlander-wbf}
 R.~Harlander, J.~Vollinga and M.~M.~Weber, in preparation.

\bibitem{DelDuca:2001fn}
  V.~Del Duca, W.~Kilgore, C.~Oleari, C.~Schmidt and D.~Zeppenfeld,
  Nucl.\ Phys.\ B {\bf 616} (2001) 367
  [arXiv:hep-ph/0108030] and
  Phys.\ Rev.\ Lett.\  {\bf 87} (2001) 122001
  [arXiv:hep-ph/0105129].


\bibitem{Denner:2003yg} A.~Denner, S.~Dittmaier, M.~Roth and
  M.~M.~Weber,
Phys.\ Lett.\ B {\bf 560} (2003) 196
[hep-ph/0301189] and
Nucl.\ Phys.\ B {\bf 660} (2003) 289
[hep-ph/0302198].

\bibitem{Hahn:2000kx}
T.~Hahn,
Comput.\ Phys.\ Commun.\  {\bf 140} (2001) 418
[hep-ph/0012260].

\bibitem{Hahn:1998yk}
T.~Hahn and M.~Perez-Victoria,
Comput.\ Phys.\ Commun.\  {\bf 118} (1999) 153
[hep-ph/9807565];
%
T.~Hahn,
Nucl.\ Phys.\ Proc.\ Suppl.\  {\bf 89} (2000) 231
[hep-ph/0005029].

\bibitem{Denner:2002ii}
A.~Denner and S.~Dittmaier,
Nucl.\ Phys.\ B {\bf 658} (2003) 175
[hep-ph/0212259].

\bibitem{Passarino:1979jh}
G.~Passarino and M.~Veltman,
Nucl.\ Phys.\ B {\bf 160} (1979) 151.

\bibitem{Denner:2005nn}
  A.~Denner and S.~Dittmaier,
  Nucl.\ Phys.\  B {\bf 734} (2006) 62
  [arXiv:hep-ph/0509141].

\bibitem{Lepage:1977sw}
G.~P.~Lepage,
J.\ Comput.\ Phys.\  {\bf 27} (1978) 192 and
CLNS-80/447.


\end{thebibliography}
\end{document}